\documentclass[cits]{PoS}
\usepackage{epsfig}

\usepackage{colordvi}
\usepackage{amsmath}   
\usepackage{amssymb}
\usepackage{fontenc}
\usepackage{times}
\usepackage{mathptmx}
\usepackage{graphicx}

\def \ar {\texttt{AMBRE}{}}
\def \mb {\texttt{MB}{}}

\newcommand{\bea}{\begin{eqnarray}}
\newcommand{\eea}{\end{eqnarray}}
\newcommand{\bqa}{\begin{eqnarray}}
\newcommand{\eqa}{\end{eqnarray}}
\newcommand{\non}{\nonumber}
\newcommand{\nl}{\nonumber \\}

\newcommand{\eps}{\varepsilon}

\newcommand{\bq}{ \begin {equation} }
\newcommand{\eq}{\end{equation}}
\newcommand{\be}{\begin{eqnarray}}
\newcommand{\ea}{\end{eqnarray}}

\newcommand{\Li}[2]{{\mathrm{Li}}_{#1}(#2)}
  \title{{\small DESY 07--104\newline SFB/CPP--07--39\newline HEPTOOLS 07-017} \\[1.5cm]
Automatizing the application of Mellin-Barnes representations
for Feynman integrals}
\ShortTitle{\ar{} -- MB-representations for Feynman integrals}

  \author{J. Gluza$^a$, F. Haas$^b$, K. Kajda$^a$ and \speaker{T. Riemann}$^b$
\\
\llap{$^a$}Department of Field Theory and Particle Physics,
    Institute of Physics, \\
    University of Silesia, Uniwersytecka 4, PL-40-007 Katowice,
    Poland
\\
\llap{$^b$}Deutsches Elektronen-Synchrotron, DESY,
   Platanenallee 6, 15738 Zeuthen, Germany
\\
Email: \email{gluza@us.edu.pl}, \email{Felix.Haas@desy.de}, \email{kkajda@us.edu.pl}, \email{Tord.Riemann@desy.de}}

\abstract{Feynman diagrams may be evaluated by Mellin-Barnes representations
of their  Feynman parameter integrals in $d=4-2\eps$ dimensions.
Recently, the Mathematica toolkit \ar{} has been developed for the automatic derivation of  such representations with a loop-by-loop approach.
We describe the package and exemplify its use with the $\eps$-expansion of the massive one-loop QED vertex function.}

\FullConference{XI International Workshop on Advanced Computing and Analysis Techniques in Physics Research, ACAT, Amsterdam, The Netherlands, April 23-27, 2007}

\begin{document}
\allowdisplaybreaks
\section{\label{secintro}Introduction}
For many of the precision  predictions of observables for LHC and ILC we need the
automated evaluation of multi-leg and/or  multi-loop Feynman diagrams; a summary of the needs and the present status may be found in \cite{Weinzierl:2007}.

A promising approach is the representation of the Feynman diagrams by Feynman parameter integrals and the subsequent use of Mellin-Barnes (MB) representations for their further evaluation.
The method was invented in 1975 for the finite massive scalar three-point function in \cite{Usyukina:1975yg}.
Massive one-loop self-energies and vertices were investigated in $d$ dimensions  in \cite{Boos:1991rg}.
An important step was the explicit evaluation, in terms of polylogarithms, of the planar on-shell double box in \cite{Smirnov:1999gc,Smirnov:1999wz,Smirnov:2001cm,Heinrich:2004iq} and of the non-planar case in \cite{Tausk:1999vh,Heinrich:2004iq}.
Slightly different algorithms are used for the derivation of proper MB-representations for divergent diagrams as expansions in powers of $\eps = (4-d)/2$.
The algorithm of Tausk was automated in Mathematica and Maple \cite{Anastasiou:2005cb} and in Mathematica \cite{Czakon:2005rk}; the latter package, \mb{}, is publicly available.
As input it needs some MB-integral representation and performs the $\eps$ expansion.
Quite recently,  the  Mathematica package \ar{} was published which prepares for a large class of Feynman diagrams the MB-integral representation \cite{Gluza:2007rt}.

In this contribution, we describe the package \ar{} and give a simple example of its use
\footnote{Examples of different complexity may be found at the webpages \cite{Katowice-CAS:2007,Zeuthen-CAS:2004}.}.

\section{\label{sec-cons}Construction of Mellin-Barnes representations}
We will use the MB-representation
\be
\frac{1}{(A+B)^{\nu}}
= \frac{B^{-\nu}}{2 \pi i\Gamma{(\nu)}}
\int\limits_{-i \infty+R}^{i \infty+R}
d \sigma {A^\sigma ~
B^{-\sigma}} ~ \Gamma{(-\sigma)}\Gamma{(\nu+\sigma)} ,
\label{mb}
\ea
where the integration contour separates the poles of the $\Gamma$-functions.

We evaluate $L$-loop Feynman integrals\footnote{Often one uses the additional normalization $e^{\eps\gamma_E L}$; we leave this to the later evaluation with the package \mb{} \cite{Czakon:2005rk}.} in $d=4-2\eps$ dimensions with $N$ internal lines with momenta $q_i$ and masses  $m_i$,  and $E$
external legs with momenta $p_e$:
\bea\label{eq-bha}
G_L[T(k)]
=
\frac{1}{(i\pi^{d/2})^L} \int \frac{d^dk_1 \ldots d^dk_L~~T(k)}
     {(q_1^2-m_1^2)^{\nu_1} \ldots (q_i^2-m_i^2)^{\nu_i} \ldots
       (q_N^2-m_N^2)^{\nu_N}  }  .
\eea
The numerator $T(k)$ is a tensor in the integration variables:
\bea\label{eq-T}
T(k) &=& 1, k_l^{\mu}, k_l^{\mu}k_n^{\nu}, \ldots 
\eea 
The momenta of the denominator functions $d_i$ are:
\bea\label{di1} 
d_i &=& q_i^2-m_i^2 ~=~
~=~ \left(\sum_{l=1}^{L} \alpha_{il} k_l -\sum_{e=1}^{E} \beta_{ie} p_e\right)^2 -m_i^2.
\eea
The momentum integrals are replaced by standard Feynman parameter integrals:
\bea
\label{eq-scalar1}
G_L[T(k)]&=& 
\frac{(-1)^{N_{\nu}} \Gamma\left(N_{\nu}-\frac{d}{2}L\right)}
{\prod_{i=1}^{N}\Gamma(\nu_i)}
\int_0^1 \prod_{j=1}^N dx_j ~ x_j^{\nu_j-1}
\delta(1-\sum_{i=1}^N x_i)
\frac{U(x)^{N_{\nu}-d(L+1)/2}}{F(x)^{N_{\nu}-dL/2}}~P_L(T),
\nl
\eea
with
\bea\label{eq-Nnu}
N_{\nu} &=& \sum_{i=1}^{N} \nu_i.
\eea
The two functions $U$ and $F$ may be derived from
\bea
\label{bh-6}
{\mathcal N} &=& \sum_{i=1}^{N} x_i (q_i^2-m_i^2) ~\equiv~ kMk - 2kQ + J,
\eea
where $M_{ll'} = \sum_{i=1}^N \alpha_{il'}\alpha_{il} x_{i} $, and $Q_l = \sum_{i=1}^N \alpha_{il} P_i x_i$, and $J = \sum_{i=1}^N (P_i^2-m_i^2)x_i$;
namely:
\bea\label{eq:U} 
U{(x)} &=& \textrm{det}( M),
\\\label{eq:F} 
F({x}) &=& 
-\textrm{det}( M)~J + Q\tilde{M}Q.
\eea
Where we may assume $M^+ = M$.
The $U$ and $F$ as well as $\tilde{M} = \textrm{det}( M) ~ M^{-1} $ are polynomials in $x$.
We will evaluate $L$-loop integrals with the
 loop-by-loop iteration procedure, because the formulae simplify for one-loop integrals:
\begin{eqnarray}
 \label{u1loop} 
U &=& M ~=~ {\tilde{M}} ~=~ \textrm{det}(M) ~=~ \sum_i^N x_i ~=~ 1,
\\ \label{f1loop} 
F &=& -U J+ Q^2 ~=~ \sum_{i,j}^N [P_iP_j-P_i^2+m_i^2]x_ix_j ~\equiv~  \sum_{i\leq j}^N f_{ij}x_ix_j .
\end{eqnarray} 
The simplest tensor factors  $P_1(T)$ in (\ref{eq-scalar1}) become:
\bea
\label{p1loop}
P_1(1) &=& 1,
\\\label{eq:p1k} 
P_1(k^\alpha) &=& \sum_{i=1}^N x_i P_i^\alpha,
\\\label{eq:p1tens2} 
P_1(k^{\alpha}k^{\beta}) &=& \sum_{i=1}^N x_i P_i^{\alpha}\sum_{j=1}^N x_j P_j^{\beta}
-
\frac{\Gamma\left(N_{\nu}-\frac{d}{2}-1\right)}
{\Gamma\left(N_{\nu}-\frac{d}{2}\right)}~F~\frac{g^{\alpha\beta}}{2} .
\eea
The $P_i^{\alpha}$ are the so-called chords introduced in (\ref{di1}).
The general case is:
\begin{eqnarray}\label{eq:tensorAnast}
G_1(T_m)&\equiv&
G_1(k^{\mu_1}\cdots k^{\mu_m})
\nl
&=&
\frac{(-1)^{N_{\nu}}}{\prod_{i=1}^N \Gamma(\nu_i)} 
\int
\prod_{i=1}^{N}dx_i x_{i}^{\nu_{i}-1} \delta(1-\sum_{j=1}^N x_j)
\sum_{r=0}^m 
\frac{\Gamma\left(N_{\nu}-\frac{d+r}{2}\right)}{ (-2)^{\frac{r}{2}} F^{N_{\nu}-\frac{d+r}{2}} }
\left\{{\cal A}_r P^{m-r}\right\}^{(\mu_1,\ldots,\mu_m)} ,
\nl
\end{eqnarray}
with the abbreviations $F\equiv F(x)$ and $P \equiv P_1(k^{\mu}) = \sum_i x_i P_i = \sum_{i,e}x_i\beta_{ie}p_e^{\mu}$.
The $r$ starts from zero (with ${\cal A}_0=1$), and it is ${\cal A}_r=0$ for $r$ odd, and
${\cal A}_r= g^{\mu_{i_1} \mu_{i_2}}\cdots g^{ \mu_{i_{r-1}} \mu_{i_r}}$
 for $r$ even. 
The convention $(\mu_{i_1} \ldots)$ means the totally symmetric combination of the arguments.
\\
In \ar{}, the  tensorial numerators are assumed to be contracted with a tensor $P(m)$ composed of external momenta $p_e$, so that the following quantity is evaluated:
\bea
P(m) ~ G_1(T_m) &\equiv& \left( p_{e_1}^{\mu_1}\cdots p_{e_m}^{\mu_m} \right)~ G_1(k^{\mu_1}\cdots k^{\mu_m}).
\eea

One now has to perform the $x$-integrations.
We do this by the following simple formula: 
\bqa
\label{a1}
\int_0^1 \prod_{i=1}^N dx_i ~ x_i^{q_i-1}
~ \delta\left(1-\sum_j x_j\right)
&=&
\frac{\Gamma(q_1) \cdots \Gamma(q_N)}
{\Gamma\left(q_1 + \cdots + q_N \right)} .
\eqa
From the above text it is evident that the integrand of  (\ref{eq-scalar1}) contains besides simple sums of monomials $\prod_i x_i^{n_i}$ (like e.g. $Q$ and $ {\tilde{M}}$) also sums of monomials with non-integer exponents.
This is due to the appearance of the factors  $U(x)$ and $F(x)$.
One may rewrite $F(x)$ and $U(x)$ so that (\ref{a1}) becomes applicable.
For the one-loop case this has to be done only for $F(x)$.
In (\ref{f1loop}), $F(x)$ is  written as  a sum of $N_F \leq \frac{1}{2}N(N+1)$ non-vanishing, bilinear terms in $x_i$\footnote{If useful, one may also consider $F(x)$ with linear and bilinear terms in the $x_i$.}:
\bea\label{fxrep} 
F(x)^{-(N_\nu-dL/2)} &=& \left[ \sum_{n=1}^{N_F} f_n(i,j) x_i x_j \right]^{-(N_\nu-dL/2)}
\nl
&=& \frac{1}{\Gamma(N_\nu-dL/2)}\frac{1}{(2\pi i)^{N_F}}
\prod_{i=1}^{N_F} \int\limits_{-i \infty + u_i}^{i \infty + u_i} d z_i \prod_{n=2}^{N_F}
\left[ f_{n}(i,j) x_i x_j\right] ^{z_n} 
\nl&&~
\left[ f_{1}(i,j)x_i x_j\right] ^{-(N_\nu-dL/2)-\sum_{j=2}^{N_F} z_j} 
\Gamma\left( N_\nu-\frac{dL}{2}+\sum_{j=2}^{N_F}z_j\right) 
\prod_{j=2}^{N_F} \Gamma(-z_j) .
\nl
\eea
Here, $f_n(i,j) = f_{ij} $ if $f_{ij} \neq 0$. 
Inserting (\ref{fxrep}) and the tensor function $P(T)$ into (\ref{eq-bha}) allows to apply (\ref{a1}) for an evaluation of the $x$-integrations. 

As a result, $L$-loop  scalar Feynman integrals may be represented by a single multi-dimensional MB-integral and tensor Feynman integrals by finite sums of MB-integrals.

\section{\label{sec:ambre}The Mathematica package \ar}
In this section we describe the use of the package \ar{} \cite{Gluza:2007rt}.  
\ar{} stands for {\bfseries{A}}utomatic {\bfseries{M}}ellin-{\bfseries{B}}arnes {\bfseries{Re}}presentation. 
It is a (semi-)automatic procedure written for multi-loop calculations. 
The package works with Mathematica 5.0 and later versions of it.
One has to perform the following tasks:
\begin{enumerate}
\item[(i)] define kinematical invariants which depend on the external momenta;
\item[(ii)] make a decision about the order in which $L$ one-loop subloops 
$(L \geq 1)$  will be worked out sequentially;
\item[(iii)] construct a Feynman integral for the chosen subloop and 
perform manipulations
on the corresponding $F$-polynomial to make it optimal for later use of the MB 
representations;
\item[(iv)] use equation (\ref{fxrep});
\item[(v)] perform the integrations over Feynman parameters with equation (\ref{a1});
\item[(vi)] if needed, go back to  step (iii) and repeat the steps for the next subloop until $F$ in 
the last, $L^{th}$ subloop will be changed into an MB-integral.
\end{enumerate}
The steps (ii) and (iii) 
must be analyzed carefully, because there exists some freedom of choice
on the order of loop integrations in step (ii) and also
on the order of MB integrations in  
step (iii). 
Different choices may lead to different forms of MB-representations.

The basic functions of \ar{} are:
\begin{itemize}
\item
{\bf Fullintegral[\{numerator\},\{propagators\},\{internal momenta\}]} -- is the basic function for 
input Feynman integrals
\item
{\bf invariants} -- is a list of invariants, e.g. {\bf invariants = \{p1*p1 $\to$ s\} }
\item
{\bf IntPart[iteration]} -- prepares a subintegral for a given internal momentum by collecting the related numerator, propagators, integration momentum
\item
{\bf Subloop[integral]} -- determines for the selected subintegral the $U$ and $F$ polynomials and an MB-representation
\item
{\bf ARint[result,i\_]} -- displays the MB-representation number i for Feynman integrals
with numerators
\item
{\bf Fauto[0]} -- allows user specified modifications of the $F$ polynomial {\bf fupc}
\item
{\bf BarnesLemma[repr,1,Shifts\verb+->+True]}
-- function tries to apply Barnes' first lemma to a given MB-representation; when {\bf Shifts\verb+->+True} is set, \ar{} will try a simplifying shift of variables; the default is  {\bf Shifts\verb+->+False} 
\\ 
{\bf BarnesLemma[repr,2,Shifts\verb+->+True]}
-- function tries to apply Barnes' second lemma
\end{itemize}

\section{\label{sec:one-loop0}V3l2m: the one-loop massive QED vertex}
As an example, we evaluate the one-loop massive QED vertex function V3l2m (vertex with 3 internal lines, 2 of them being massive, $m^2=1$):
\bea
{\texttt{V3l2m}} =  e^{\eps\gamma_E} \int \frac{d^{d}k}{i\pi^{d/2}}
 \frac{1}{(k^2)^{n_{0}}[(k+p_{1})^2-1]^{n_{1}}[(k-p_{2})^2-1]^{n_{2}}}.
\eea
The corresponding definition in \ar{}:

{\bf >> V3l2m = Fullintegral[\{1\},\{PR[k, 0, n0] PR[k + p1, 1, n1] PR[k - p2, 1, n2]\},\{k\}] }
\\
In a next step we define the invariants:

{\bfseries
>> invariants = $\{p1^2 -> 1, p2^2 -> 1, p1 p2 -> (s - 2)/2\}$}
\\
With {\bfseries IntPart[1]} and {\bfseries SubLoop[integral]}, we determine the $F$-polynomial for the diagram,
\bea
F=(x_1+x_2)^2 +[- s] x_1 x_2,
\eea
As usual, we have to replace here a positive $s$ by $s+i\varepsilon$.
After setting  {\bfseries \{n1 -> 1, n2 -> 1, n0 -> 1\}} and applying Barnes' first lemma,

{ \bf >> MBV3l2m = BarnesLemma[V3l2m, 1]}
\\
an MB-representation for the Feynman integral is obtained:
\bea
\label{defv3l}
 {\tt V3l2m[y]} &=& 
-\frac{e^{\eps\gamma_E}\Gamma(-2\eps)}{\Gamma(1-2\eps)2\pi i}
\int dz (-s)^{-\eps-1-z}
\frac{\Gamma^2(-\eps-z)\Gamma(-z)\Gamma(1+\eps+z)} {\Gamma(-2\eps-2z)} .
\eea
For the subsequent step, the derivation of the $\eps$-expansion, the package \mb{} may be used:
\bea
{\tt V3l2m[y]} &=& \frac{ {\tt V3l2m[-1,y]} }{\eps} + {\tt V3l2m[0,y]}
+ \eps~{\tt V3l2m[1,y]}+\cdots,
\eea
where we already introduced the conformal variable
\bea
y=\frac{\sqrt{-s+4}-\sqrt{-s}}{\sqrt{-s+4}+\sqrt{-s}}.
\eea
Representations for the first terms of the $\eps$-expansion are easily obtained with the following \mb{} commands

{ \bf >> rules = MBoptimizedRules[MBV3l2m, eps -> 0, \{\}, \{eps\}]}

{ \bf >> integrals = MBcontinue[MBV3l2m, eps -> 0, rules]}

{ \bf >> expV3l2m = MBexpand[integrals, Exp[eps*EulerGamma], \{eps, 0, n\}]}
\\
We reproduce a few of them here:
\bea
{\tt V3l2m[-1,y]} 
&=&  
\frac{1}{2}\frac{1}{2\pi i}\int_{-i\infty+u}^{+i\infty +u}dz
(-s)^{-1-z}
~\frac{\Gamma^3[-z]\Gamma[1 + z]}{ \Gamma[-2z]},
\\
{\tt V3l2m[0,y]} &=& 
\frac{1}{2\pi i}\int_{-i\infty+u}^{+i\infty +u}dz
(-s)^{-1-z}
~\frac{\Gamma^3[-z]\Gamma[1 + z]}{ \Gamma[-2z]}
\nl&&
\frac{1}{2}\left[ \gamma_E - \ln(-s) + 2 \Psi[-2z] -2\Psi[-z] + \Psi[1+z]\right],
\\
{\tt V3l2m[1,y]} &=&  
\frac{1/4}{2\pi i}\int_{-i\infty+u}^{+i\infty +u}dz
(-s)^{-1-z}
~\frac{\Gamma^3[-z]\Gamma[1 + z]}{ \Gamma[-2z]}
\nl&&
\Bigl[ 
\gamma_E^2 + Log[-s]^2
+
 Log[-s](-2\gamma_E - 4 \Psi[ -2z] + 4 \Psi[ -z] -  2 \Psi[ 1 + z])
\nl &&
+\gamma_E (4  \Psi[ -2 z] - 4  \Psi[ -z] + 
        2  \Psi[ 1 + z]) 
\nl &&
- 4  \Psi[1, -2 z] + 2  \Psi[1, -z] + 
   \Psi[1, 1 + z]
\nl &&
+ 4( \Psi[ -2z]^2 - 2 \Psi[ -2z] \Psi[ -z] + 
  \Psi[ -z]^2 +  \Psi[ -2z] \Psi[ 1 + z] 
\nl &&
- 
  \Psi[ -z] \Psi[ 1 + z]) +  \Psi[ 1 + z]^2 
\Bigr],
\\
{\tt V3l2m[2,y]} &=&  
\frac{1/12}{2\pi i}\int_{-i\infty+u}^{+i\infty +u}dz
(-s)^{-1-z}
~\frac{\Gamma^3[-z]\Gamma[1 + z]}{ \Gamma[-2z]}
\nl&&
\Bigl[ {a(z)}+{b(z)} \Psi(0,1+z)+ \Psi(2,1+z) +2\Psi(0,1+z)^2+\Psi(0,1+z)^3
\nl&&
+~3\Psi(0,1+z)\Psi(1,1+z)
+{c(z)} [\Psi(1,1+z)+2\Psi(0,1+z)^2],
\Bigr]
\eea
with some coefficients $a(z), b(z), c(z) $, which depend on $ Log[-s], \gamma_E, \Psi(k,-z), \Psi(k,-2z) $ ($k=0,1,2$).
Here, $\Psi[z]$ and $\Psi[k, z]$ are polygamma functions:
\bea  
\textrm{PolyGamma [z]} &=& \textrm{PolyGamma [0,z]}
~=~\Psi(z)
~=~\frac{\Gamma'(z)}{\Gamma(z)}
\\
\non
\textrm{PolyGamma [0,n+1]} 
&=& 
~S_1(n) - \gamma_E
~=~\sum_{i=1}^{n}\frac{1}{i} - \gamma_E,
\\
\textrm{PolyGamma [n,z]} &=&\Psi[n, z] =  \frac{d^n\Psi(z)}{dz^n} = (-1)^{n+1}n!\sum_{k=0}^{\infty}\frac{1}{(z+k)^{n+1}}.
\eea
The integration path includes the straight line ranging from $(-i\infty+1/2)$ to $(+i\infty+1/2)$.
Closing the path to the left, allows to express the integral as an infinite series of residua arising from the poles of
\bea
\Gamma[1 + z], \Psi[k,1+z],
\eea
at $z = -n, n = 1, 2, \cdots$  with different weight functions $G(z)$, e.g. for {\tt V3l2m[-1,y]} with
\bea
G(z) = (-s)^{-1-z}~\frac{\Gamma^3[-z]}{ \Gamma[-2z]}.
\eea
The resulting respresentations are inverse binomial sums:
\bea\label{v3l2mm1}
{\tt V3l2m[-1,y]} &=&  \frac{1}{2}\sum_{n=0}^{\infty}  \frac{s^n} { \binom{2n}{n} (2n+1)},
\\\label{v3l2m0}
{\tt V3l2m[0,y]} &=&  \frac{1}{2}\sum_{n=0}^{\infty}  \frac {s^n} { \binom{2n}{n} (2n+1)} S_1(n),
\\\label{v3l21}
{\tt V3l2m[1,y]} &=&  
\frac{1}{4} \sum_{n=0}^{\infty}  \frac{s^n} { \binom{2n}{n} (2n+1)}
\left[ S_1(n)^2 +\zeta_2 - S_2(n)\right],
\\\label{v3l22}
{\tt V3l2m[2,y]} &=& \sum_{n=0}^{\infty}  \frac{s^n} { \binom{2n}{n} (2n+1)}
\Bigl[
\frac{1}{12}S_1[n]^3-\frac{1}{4}S_1[n]S_2[n]
  +\frac{1}{4}\zeta_2  S_1[n]
+\frac{1}{6}S_3[n]
-\frac{1}{6}\zeta_3\Bigr],
\nonumber\\
\eea
etc.,
and the harmonic numbers $S_k(n)$ are
\bea
\textrm{HarmonicNumber[n,k]} ~=~ S_k(n) &=& \sum_{i=1}^{n}\frac{1}{i^k}.
\eea
The simplest of the sums may be done  with Mathematica:
\bea
{\tt V3l2m[-1,y]} 
&=& 
\frac{1}{2}\frac{4\arcsin(\sqrt{s}/2)}{\sqrt{4-s}\sqrt{s}}
~=~\frac{1}{2}\frac{-2y}{1-y^2}\ln y.
\eea
The other sums appearing  above may be obtained from sums listed in Table 1 of Appendix D in \cite{Davydychev:2003mv}:
\begin{eqnarray}
\sum_{n=0}^\infty \frac{s^n}{\binom{2n}{n} (2n+1)}&=&\frac{y}{y^2-1} 2\ln (y),
\\
\sum_{n=0}^\infty \frac{s^n}{\binom{2n}{n}   (2n+1)}S_1(n)&=&\frac{y}{y^2-1}
\left[ -4 \Li{2}{-y} - 4 \ln (y) \ln (1+y)\right.
+\left.\ln^2 (y) - 2\zeta_2 \right],
\label{S1_y}
\\
\sum_{n=0}^\infty \frac{s^n}{\binom{2n}{n}   (2n+1)}S_1(n)^2&=&\frac{y}{y^2-1}
\biggl[
16 S_{1,2}(-y)
- 8 \Li{3}{-y}
+ 16 \Li{2}{-y} \ln(1+y)
\nonumber \\ &&
+ 8 \ln^2 (1+y) \ln (y)
- 4 \ln(1+y) \ln^2 (y)
+ \frac{1}{3} \ln^3 (y)
+ 8 \zeta_2 \ln(1+y)
\nonumber \\ &&
- 4 \zeta_2 \ln (y)
- 8 \zeta_3
\biggr],
\\
\sum_{n=0}^\infty \frac{s^n}{\binom{2n}{n}   (2n+1)}S_2(n)&=&-\frac{y}{3(y^2-1)}\ln^3(y),
\\
\sum_{n=0}^\infty \frac{s^n}{\binom{2n}{n}   (2n+1)}S_1(n)^3&=&\frac{y}{y^2-1}
\Bigl[
- 96 S_{1,2}(-y) \ln (1+y)
- 96 S_{1,3}(-y)
+ 48 S_{2,2}(-y)
\nonumber \\ &&
- 24 \zeta_2 \ln^2 (1+y)
- 48 \ln^2 (1+y) \Li{2}{-y}
+ 48 \zeta_3 \ln (1+y)
\nonumber \\ &&
+ 48 \ln(1+y) \Li{3}{-y}
- 16 \ln (y) \ln^3 (1+y)
+ 24 \zeta_2 \ln (y) \ln (1+y)
\nonumber \\ &&
+ 12 \ln^2 (y) \ln^2 (1+y)
- 2\ln^3 (y) \ln(1+y)
+ \frac{1}{12} \ln^4 (y)
- 3 \zeta_2 \ln^2 (y)
\nonumber \\ &&
+ 6 \ln^2 (y) \Li{2}{-y}
+ 2\ln^2 (y) \Li{2}{y}
 - 10 \zeta_3 \ln (y)
- 24 \ln (y) \Li{3}{-y}
\nonumber \\ &&
- 8 \ln (y) \Li{3}{y}
+ 3 \zeta_4
+ 24 \Li{4}{-y}
+ 12 \Li{4}{y}
\Bigr],
\\
\sum_{n=0}^\infty \frac{s^n}{\binom{2n}{n}   (2n+1)}S_3(n)&=&\frac{y}{y^2-1}
\Bigl[
\frac{1}{12} \ln^4 (y)
+ 12 \Li{4}{y}
+ 2\ln^2 (y) \Li{2}{y}
\nonumber \\ &&
- 4 \zeta_3 \ln (y)
- 8 \ln (y) \Li{3}{y}
 - 12 \zeta_4
\Bigr],
\\
\sum_{n=0}^\infty \frac{s^n}{\binom{2n}{n}   (2n+1)}S_1(n)S_2(n)&=&\frac{y}{y^2-1}\Bigl[
\frac{2}{3} \ln^3 (y) \ln(1\!+\!y)
- \! \frac{1}{12} \ln^4 (y)
+ \!  \zeta_2 \ln^2 (y)
\nonumber \\ &&
+ 2\ln^2 (y) \Li{2}{-y}
+ 2\ln^2 (y) \Li{2}{y}
+ 2\zeta_3 \ln (y)
\!-\! 8 \ln (y) \Li{3}{-y}
\nonumber \\ &&
\!-\! 8 \ln (y) \Li{3}{y}
\!+\! 2\zeta_4
+ \! 16 \Li{4}{-y}
+ \! 12 \Li{4}{y}
\Bigr].
\end{eqnarray}
The one-loop QED vertex function is a relatively  simple one-scale problem, and in fact one may derive the general term of the $\eps$-expansion.
So far, we applied the Mathematica packages without an interference by the user.
The packages have been created for the automatization of this type of calculations.
But here we have an instructive example of the limitations of that.
Let us go back to Equation (\ref{defv3l}) and shift the integration variable $z$ according to $z = z'-\eps$, with a related shift of the real part of the integration path, $u'=u+\eps$.
The resulting MB-integral is:
\bea
\label{defv3la}
 {\tt V3l2m[y]} &=& 
-\frac{e^{\eps\gamma_E}\Gamma(-2\eps)}{\Gamma(1-2\eps)2\pi i}\int dz' (-s)^{-1-z'}\frac{\Gamma^2(-z')\Gamma(-z'+\eps)\Gamma(1+z')}
{\Gamma(-2z')}
\eea
After taking residua as before, but now without using \mb{}, one has to evaluate:
\bea
 {\tt V3l2m[y]} &=& 
\frac {e^{\eps\gamma_E}} {2\eps} \sum_{n=0}^{\infty} \frac{s^n}{\binom{2n}{n}(2n+1)} 
\frac{\Gamma(n+1+\eps)}{\Gamma(n+1)}.
\eea
We may apply here the following well-known relation \cite{Weinzierl:2004bn}\footnote{Express $\Gamma(-n+\eps)$ by $\Gamma(\eps)$.
Thereby factors $1/(-n+\eps+k)$ are collected,  inverse them for small $\eps$.}:
\bea
\frac{\Gamma(n+1+\eps)}{\Gamma(n+1)} &=& \Gamma(1+\eps) 
\exp \left[ -\sum_{k=1}^{\infty} \frac{(-\eps)^k}{k} S_k(n)\right],
\eea
and obtain the general term of the inverse binomial sum for the $n^{th}$ term in the $\eps$-expansion of {\tt V3l2m[y]}.
The first terms agree with (\ref{v3l2mm1})--(\ref{v3l22}).
The evaluation of the sums 
may be performed for the general case by introducing integral representations for the $\Gamma$ functions and the harmonic numbers (for the latter see e.g. \cite{Blumlein:1998if}), and then performing the two-dimensional integral.

Alternatively to the approach described here, one may determine the vertex function in terms of Harmonic Polylogarithms (HPLs) \cite{Remiddi:1999ew} by the differential equation method, which allows easily to derive the $\eps$-expansion of the Feynman integral 
\texttt{SE2l2m}. 
The vertex may be related to \texttt{SE2l2m} and to the tadpole \texttt{T1l1m} by integration by parts; see e.g. \cite{Fleischer:2006ht}:
\begin{eqnarray}
{\tt V3l2m}  = \frac{(d-2) {\tt T1l1m} - 2(d-3){\tt SE2l2m}}{(d-4)(s-4)}.
\end{eqnarray}
The master integrals  {\tt T1l1m} and {\tt SE2l2m} are tabulated e.g. in the file \texttt{mastersHPL.m} \cite{Czakon:2004wm,webPage:2007xx}.
From the corresponding formulae it may be seen that only HPLs with arguments 0 and -1 appear.
An approach to describe the vertex in $d$ dimensions with hypergeometric functions and the subsequent derivation of the $\eps$-expansion may be found in \cite{Davydychev:2000na}.
\section{Summary} 
To summarize, for many applications of present phenomenological or more theoretical interest
the package \ar{} solves an important  part of the  calculational problem: 
the semi-automatic derivation of MB-representations for a large class of Feynman integrals.

\acknowledgments{We would like to thank J. Bl{\"u}mlein, S.-O. Moch and B. Tausk for useful discussions and suggestions.
The present work is supported in part 
by the European Community's Marie-Curie Research Training Networks  MRTN-CT-2006-035505 `HEPTOOLS'
and MRTN-CT-2006-035482 `FLAVIAnet', 
and by  
Sonderforschungsbe\-reich/Trans\-regio 9--03 of Deutsche Forschungsgemeinschaft
`Computergest{\"u}tzte Theo\-re\-ti\-sche Teil\-chen\-phy\-sik'. }

\providecommand{\href}[2]{#2}\begingroup\raggedright\endgroup


\end{document}